\documentclass[a4,aps,twocolumn,showpacs,floatfix]{revtex4}
\usepackage{graphicx}
\usepackage{dcolumn}
\usepackage{color}
\usepackage{amsmath}
\usepackage{here}

\begin{document}

\title{Tunneling of persistent currents in coupled ring-shaped Bose-Einstein
condensates}
\author{ Artem Oliinyk$^{1}$, Alexander Yakimenko$^{1}$, Boris Malomed$^2$ }
\affiliation{$^1$ Department of Physics, Taras Shevchenko National University of Kyiv,
64/13, Volodymyrska Street, Kyiv 01601, Ukraine \\
$^2$ Department of Physical Electronics, Faculty of Engineering, and Center
for Light-Matter Interaction, Tel Aviv University, Tel Aviv 69978, Israel}

\begin{abstract}
Considerable progress in experimental studies of atomic gases in a toroidal
geometry opens up novel prospects for the investigation of fundamental
properties of superfluid states and creation of new configurations for
atomtronic circuits. In particular, a setting with Bose-Einstein condensates
loaded in a dual-ring trap suggests a possibility to consider the dynamics
of tunneling between condensates with \emph{different} angular momenta.
Accordingly, we address the tunneling in a pair of coaxial three-dimensional
(3D) ring-shaped condensates separated by a horizontal potential barrier. A
truncated (finite-mode) Galerkin model and direct simulations of the
underlying 3D Gross-Pitaevskii equation are used for the analysis of
tunneling superflows driven by an initial imbalance in atomic populations of
the rings. The superflows through the corresponding Bose-Josephson junction
are strongly affected by persistent currents in the parallel rings.
Josephson oscillations of the population imbalance and angular momenta in
the rings are obtained for co-rotating states and non-rotating ones. On the
other hand, the azimuthal structure of the tunneling flow demonstrates
formation of Josephson vortices (fluxons) with zero net current through the
junction for \textit{hybrid states}, built of \textit{counter-rotating
persistent currents} in the coupled rings.
\end{abstract}

\pacs{}
\maketitle

\section{Introduction}

The Josephson effect in alternating- and direct-current (a.c. and d.c.)
forms was first discovered in a pair of two superconductors, the macroscopic
wave functions of which are weakly coupled across the tunneling barrier \cite%
{Josephson62, Barone82}. The a.c. Josephson effect in atomic Bose-Einstein
condensates (BEC) was experimentally observed \cite{Levy07} in an atomic
cloud loaded in the double-well potential, where a constant
chemical-potential difference between the coupled condensates drives an
oscillating atomic flow through the barrier, see also recent experimental
work \cite{Schmiedmayer}. Josephson junctions are also known in excitonic
condensates in bilayered materials \cite{bilayer1}

On the other hand, persistent currents in toroidal atomic BECs were
extensively investigated, both experimentally and theoretically, as a
hallmark of superfluidity. The ring-shaped trap produces a large central
hole around the axis of the condensate cloud. Thus, the core of persistent
currents, alias vortex states, in toroidal traps is bounded by the potential
structure, which makes even multicharged vortices robust. Further, the
stability of the persistent current in a single ring suggests one to
consider the impact of quantized angular momenta in two parallel-coupled
superfluid rings on the Josephson effect in the dual-ring setting. In
superconductors, geometrically similar ring-shaped long Josephson junctions
are well-known objects \cite{J-ring1,J-ring6,J-ring2,J-ring3, J-ring4,
J-ring7, J-ring8, J-ring5, Sherill79,Sherill81, Tilley66}, including their
discrete version \cite{discrete1,discrete2}.

\begin{figure}[h]
\centering
\includegraphics[width=3.4in]{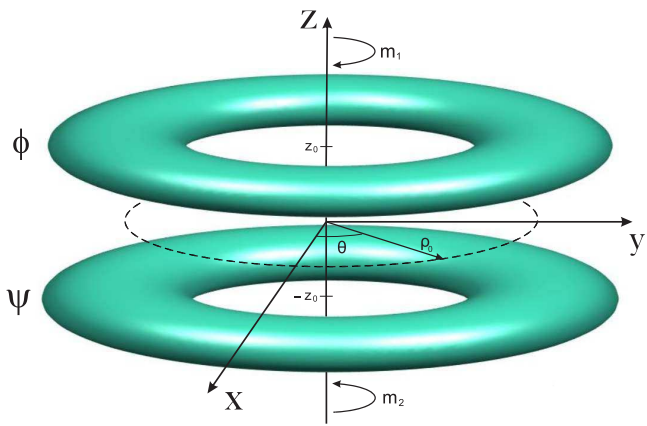}
\caption{(Color online) The schematic geometry of two weakly coupled coaxial
ring-shaped condensates, with, generally speaking, different vorticities $%
m_{1}$ and $m_{2}$ of the two components. Shown are isosurfaces of
condensate density distributions. }
\label{BECRings}
\end{figure}

Previous theoretical studies \cite{Lesanovsky07, Brand10, Zhang13,
Polo16,Brand09, Brand18, Luigi14, Davit13,
Haug18,PhysRevA.96.013620,Gallemi16} have drawn considerable interest to
systems of coupled circular BECs. Two identical parallel coaxial BEC rings,
separated in the axial direction by a potential barrier, were considered in
the context of the spontaneous generation of vortex lines \cite{Montgomery10}
and defects by means of the Kibble-Zurek mechanism \cite{Su13}. However,
Josephson dynamics in such a symmetric double-ring system, to the best of
our knowledge, has not been previously investigated. In the present work, we
address tunneling of weakly coupled quantized superflows in a pair of
stacked ring-shaped condensates. The schematic of the double-ring geometry
is displayed in Fig. \ref{BECRings}. We perform the analysis, in parallel,
on the basis of the full three-dimensional (3D) Gross-Pitaevskii equation
(GPE) for this setting and a simple finite-mode truncation, i.e., the
Galerkin approximation (GA). In the general form, the GA is introduced with
six degrees of freedom, which represent angular modes in the
parallel-coupled rings with azimuthal quantum numbers (vorticities) $0$ and $%
\pm 1$. The GA is introduced and investigated in Section II. For inputs with
identical vorticities in the coupled rings, the GA predicts a.c. Josephson
oscillations. Other remarkable dynamical states, which are produced by the
numerical solution of the full GPE, are \textit{hybrids} composed of modes with opposite
vorticities, $\left( -1,1\right) $, in the top and bottom rings. They
generate an azimuthally periodic pattern of the tunneling superflow current
along the rings, with zero net tunneling rate (no spatially average a.c.
Josephson effect). Similar conclusions are obtained for \textit{semi-vortex
hybrids} with the vorticity set of the $\left( 1,0\right) $ type
(semi-vortices, also known as half-vortices \cite{Han-Pu}, are natural modes
in spin-orbit-coupled systems \cite{Ben-Li}). Section III reports results of
systematic simulations of the full GPE, which well corroborate the GA
predictions. Thus, the six-mode GA identifies the minimal set of modes which
adequately captures basic features of the two-coupled-rings system. Beyond
the framework of the six-mode GA, the 3D simulations demonstrate that the
vanishing of the net tunneling superflow in the hybrid configurations is
associated with generation of Josephson vortices (fluxons) trapped in the
Bose Josephson junction (cf. Ref. \cite{Brand09}), which is also shown in
Section III. The paper is concluded by Section IV.

\section{The Galerkin approximation (GA)}

%%%%%%%%%%%%%%%%%%%%%%%%%%%%%%%%%%%%%%%%%
The basic set of modes which determines tunneling in the system of two
weakly parallel-coupled rings can be identified by means of a finite-mode
(truncated) approximation, i.e., GA, which replaces the underlying 3D GPE by
a dynamical system with several degrees of freedom. In this section, we
introduce the GPE for the present model, which is followed by the derivation
of the GA and analysis of major scenarios for the dynamics of imbalance of
the populations and angular momenta in the coupled annular-shaped BECs in
the framework of this approximation.

\subsection{Basic equations}

The underlying GPE is \cite{Pit}:
\begin{equation}
i\hbar \partial _{\tau }\Psi =-\frac{\hbar ^{2}}{2M}\nabla ^{2}\Psi +V_{%
\mathrm{ext}}(\mathbf{r})\Psi +g|\Psi |^{2}\Psi ,  \label{GPE3D}
\end{equation}%
where $\tau $ is the temporal variable, $\nabla ^{2}$ acts on the 3D wave
function, $g=4\pi \hbar ^{2}a_{s}/M$ is the nonlinearity strength, while $%
M=3.819\times 10^{-26}$ kg is the atomic mass and $a_{s}=2.75$ nm the $s$%
-wave scattering length for the condensate composed of $^{23}$Na atoms. In
cylindrical coordinates $\left( \rho \equiv \sqrt{x^{2}+y^{2}},\theta
,z\right) $, the trapping potential includes terms which provide radial
confinement centered at $\rho =\rho _{0}$ and a symmetric double-well
potential in the vertical direction, with minima at $z=\pm z_{0}$:
\begin{equation}
V_{\mathrm{ext}}(\rho ,z)=\frac{1}{2}M\omega _{r}^{2}(\rho -\rho _{0})^{2}+%
\frac{1}{2}M\omega _{z}^{2}z^{2}+U_{b}e^{-\frac{z^{2}}{2a^{2}}},
\label{Vext}
\end{equation}%
with $z_{0}=a\sqrt{2\ln \left[ U_{b}/(Ma^{2}\omega _{z}^{2})\right] }$.

The 3D wave function of the double-ring system can be approximately written
in terms of effective 1D wave functions of the two rings, $\phi (\theta )$
and $\psi (\theta )$, as
\begin{equation}
\Psi (\mathbf{r})=\Psi _{\rho }(\rho )\left[ \Phi _{+}(z)\frac{\phi (\theta )%
}{\sqrt{\gamma }}+\Phi _{-}(z)\frac{\psi (\theta )}{\sqrt{\gamma }}\right] ,
\label{3}
\end{equation}%
with appropriate wave functions $\Psi (\rho )$ and $\Phi _{\pm }(z)=\Phi
(z\mp z_{0})$, centered, respectively, at $\rho =\rho _{0}$ and $z=\pm z_{0}$%
, and coefficients
\begin{equation}
\gamma =8\pi a_{s}R^{2}\int_{0}^{+\infty }|\Psi _{\rho }|^{4}\rho d\rho
\int_{-\infty }^{+\infty }|\Phi (z)|^{4}dz,
\end{equation}%
\begin{equation}
R^{-2}=\int_{0}^{+\infty }|d\Psi /d\rho |^{2}\rho ^{-2}d\rho .
\end{equation}%
The first and second terms in Eq. (\ref{3}) pertain to the top and bottom
rings, respectively.

In the scaled form, the corresponding GPE system for the azimuthal wave
functions of the upper and lower tunnel-coupled rings, $\phi (\theta ,t)$
and $\psi (\theta ,t)$, produced by the substitution of ansatz (\ref{3}) in
Eq. (\ref{GPE3D}) and averaging in the directions of $\rho $ and $z$, takes
the form of \cite{Lesanovsky07, Brand10}
\begin{eqnarray}
i\phi _{t} &=&-\frac{1}{2}\phi _{\theta \theta }+|\phi |^{2}\phi -\kappa
\psi ,  \notag \\
&&  \label{coupled} \\
i\psi _{t} &=&-\frac{1}{2}\psi _{\theta \theta }+|\psi |^{2}\psi -\kappa
\phi ,  \notag
\end{eqnarray}%
where $t=\left( \hbar /2MR^{2}\right) \tau $ is dimensionless time, and
\begin{equation}
\kappa =\left\vert R^{2}\int_{-\infty }^{+\infty }\Phi _{-}^{\ast
}(z)(\partial _{z}^{2}-\frac{2M}{\hbar ^{2}}V_{z})\Phi _{+}(z)dz\right\vert .
\end{equation}%
In Eq. (\ref{coupled}), we keep coupling constant $\kappa $ unscaled, to
consider weak- and strong-coupling regimes separately. The sign of the
nonlinear terms in Eq. (\ref{coupled}) is self-repulsive, as we aim to focus
on the case when the dynamics of persistent currents in the ring-shaped BEC
is not subject to the modulational instability in the azimuthal direction,
occurs in the case of self-attraction. In the conservative system the total
number of particles and angular momentum are conserved: $N=N_{1}+N_{2}=$%
const, $L=L_{1}+L_{2}=$const, where these quantities for the upper and lower
rings are
\begin{eqnarray}
N_{1} &=&\int\limits_{0}^{2\pi }|\phi (\theta )|^{2}d\theta
,N_{2}=\int\limits_{0}^{2\pi }|\psi (\theta )|^{2}d\theta ,  \notag \\
&&  \label{NL} \\
L_{1} &=&-i\int\limits_{0}^{2\pi }\phi ^{\ast }\frac{\partial \phi }{%
\partial \theta }d\theta ,L_{2}=-i\int\limits_{0}^{2\pi }\psi ^{\ast }\frac{%
\partial \psi }{\partial \theta }d\theta .  \notag
\end{eqnarray}

It is relevant to compare Eqs. (\ref{coupled}) with similar equations which
do not include the linear coupling ($\kappa =0$), but feature nonlinear
interaction between $\phi $ and $\psi $, represent a two-component wave
function in the 1D ring, or approximate a 2D annular region \cite{Viskol}.
In that case, a relevant problem is the miscibility-immiscibility transition
in the binary condensate.

The configuration admitting counter-circulating flows, characterized by
vorticities (topological charges) $m_{1,2}=\pm m$, see Fig. \ref{BECRings},
may be approximated by the following finite-mode ansatz, which takes into
account the possible presence of the non-rotating component too:
\begin{eqnarray}
\phi \left( \theta ,t\right) &=&a_{0}(t)+a_{+}(t)e^{im\theta
}+a_{-}(t)e^{-im\theta },  \notag \\
&&  \label{ab} \\
\psi \left( \theta ,t\right) &=&b_{0}(t)+b_{+}(t)e^{im\theta
}+b_{-}(t)e^{-im\theta }.  \notag  \label{2h4}
\end{eqnarray}%
Effective evolution equations for amplitudes $a_{\pm }$ and $b_{\pm }$ can
be derived from the Lagrangian of Eq. (\ref{coupled}):
\begin{eqnarray}
\Lambda &=&\int_{0}^{2\pi }\left\{ \left[ \left( \frac{i}{2}\left( \phi
^{\ast }\phi _{t}+\psi ^{\ast }\psi _{t}\right) +\kappa \phi ^{\ast }\psi
\right) +\mathrm{c.c.}\right] \right.  \notag \\
&&\left. -\frac{1}{2}\left( \left\vert \phi _{\theta }\right\vert
^{2}+\left\vert \psi _{\theta }\right\vert ^{2}\right) -\frac{1}{2}\left(
|\phi |^{4}+|\psi |^{4}\right) \right\} d\theta .  \label{L}
\end{eqnarray}%
The substitution of ansatz (\ref{ab})\ in this expression and integration
yields%
\begin{eqnarray}
\frac{\Lambda }{2\pi } &=&\frac{i}{2}\sum_{+,-}\left( a_{0}^{\ast }\frac{%
da_{0}}{dt}+b_{0}^{\ast }\frac{db_{0}}{dt}+a_{\pm }^{\ast }\frac{da_{\pm }}{%
dt}+b_{\pm }^{\ast }\frac{db_{\pm }}{dt}\right)  \notag \\
&&+\mathrm{c.c.}-H,  \label{Lagr}
\end{eqnarray}%
where $\mathrm{c.c.}$stands for the complex conjugate expression, and the
Hamiltonian is%
\begin{gather}
H=\frac{m^{2}}{2}\left( \left\vert a_{+}\right\vert ^{2}+\left\vert
a_{-}\right\vert ^{2}+\left\vert b_{+}\right\vert ^{2}+\left\vert
b_{-}\right\vert ^{2}\right)  \notag \\
-\left[ \kappa \left( a_{0}b_{0}^{\ast }+a_{+}b_{+}^{\ast }+a_{-}b_{-}^{\ast
}\right) +\mathrm{c.c.}\right]  \notag \\
+\frac{1}{2}\left[ \left\vert a_{0}\right\vert ^{4}+\left\vert
a_{+}\right\vert ^{4}+\left\vert a_{-}\right\vert ^{4}+4\left\vert
a_{+}\right\vert ^{2}\left\vert a_{-}\right\vert ^{2}\right.  \notag \\
\left. +\left\vert b_{+}\right\vert ^{4}+\left\vert b_{-}\right\vert
^{4}+4\left\vert b_{+}\right\vert ^{2}\left\vert b_{-}\right\vert ^{2}\right.
\notag \\
+2\left( a_{0}^{2}a_{+}^{\ast }a_{-}^{\ast }+b_{0}^{2}b_{+}^{\ast
}b_{-}^{\ast }+\mathrm{c.c.}\right)  \notag \\
\left. +4\left\vert a_{0}\right\vert ^{2}\left( \left\vert a_{+}\right\vert
^{2}+\left\vert a_{-}\right\vert ^{2}\right) +4\left\vert b_{0}\right\vert
^{2}\left( \left\vert b_{+}\right\vert ^{2}+\left\vert b_{-}\right\vert
^{2}\right) \right] ,  \label{H}
\end{gather}%
cf. Ref. \cite{Lesanovsky07}. The effective Lagrangian (\ref{Lagr})\ gives
rise to the system of dynamical Euler-Lagrange equations:%
\begin{eqnarray}
i\frac{da_{0}}{dt}+\kappa b_{0} &=&H_{a0}a_{0}+2a_{+}a_{-}a_{0}^{\ast },
\notag \\
i\frac{db_{0}}{dt}+\kappa a_{0} &=&H_{b0}b_{0}+2b_{+}b_{-}b_{0}^{\ast }
\notag \\
i\frac{da_{+}}{dt}+\kappa b_{+} &=&H_{a+}a_{+}+a_{0}^{2}a_{-}^{\ast },
\notag \\
i\frac{da_{-}}{dt}+\kappa b_{-} &=&H_{a-}a_{-}+a_{0}^{2}a_{+}^{\ast },
\notag \\
i\frac{db_{+}}{dt}+\kappa a_{+} &=&H_{b+}b_{+}+b_{0}^{2}b_{-}^{\ast },
\notag \\
i\frac{db_{-}}{dt}+\kappa a_{-} &=&H_{b-}b_{-}+b_{0}^{2}b_{+}^{\ast },
\label{GalerkinSet}
\end{eqnarray}%
where $H_{a0}=\left\vert a_{0}\right\vert ^{2}+2\left\vert a_{+}\right\vert
^{2}+2\left\vert a_{-}\right\vert ^{2}$, $H_{b0}=\left\vert b_{0}\right\vert
^{2}+2\left\vert b_{+}\right\vert ^{2}+2\left\vert b_{-}\right\vert ^{2}$, $%
H_{a\pm }=m^{2}/2+\left\vert a_{\pm }\right\vert ^{2}+2\left\vert a_{\mp
}\right\vert ^{2}+2\left\vert a_{0}\right\vert ^{2}$, $H_{b\pm
}=m^{2}/2+\left\vert b_{\pm }\right\vert ^{2}+2\left\vert b_{\mp
}\right\vert ^{2}+2\left\vert b_{0}\right\vert ^{2}$ %
%It is relevant to stress that these particular sets of amplitudes must correspond to
%invariant subsystems of system (\ref{GalerkinSet}).
. This system of six evolution equations represents the GA, i.e., the
finite-mode truncation replacing the full GPE system. It admits an invariant
reduction to four equations, by setting $a_{0}=b_{0}=0$. The GA provides an
adequate simplification for diverse nonlinear systems \cite{Galerkin1}-\cite%
{Galerkin3}, including GPE-based models of trapped BEC \cite%
{Paderborn,Elad,Newcastle}.

Further, the substitution of ansatz (\ref{ab}) in the definitions of the
total number of particles and angular momentum, given by Eq. (\ref{NL}),
yields the following expressions for the GA versions of these dynamical
invariants:

\begin{gather}
N/\left( 2\pi \right) =\left\vert a_{0}\right\vert ^{2}+\left\vert
b_{0}\right\vert ^{2}+\sum_{\pm }\left( \left\vert a_{\pm }\right\vert
^{2}+\left\vert b_{\pm }\right\vert ^{2}\right) ,  \notag \\
L/\left( 2\pi \right) =\pm m\sum_{\pm }\left( \left\vert a_{\pm }\right\vert
^{2}+\left\vert b_{\pm }\right\vert ^{2}\right) .  \label{integr}
\end{gather}%
In the general case, system (\ref{GalerkinSet}) is equivalent to the
Hamiltonian one with six degrees of freedom. The presence of only three
dynamical invariants, represented by Hamiltonian (\ref{H}) and the norm and
angular momentum, given by Eq. (\ref{integr}), suggests that the GA system
is not integrable, in agreement with the well-known fact that the system of
coupled GPEs (\ref{coupled}) is a non-integrable one. The invariant
reduction of Eq. (\ref{GalerkinSet}) to four degrees of freedom, produced by
setting $a_{0}=b_{0}=0$, is not integrable either.

\subsection{Analysis of the Galerkin approximation}

Equations (\ref{GalerkinSet}), produced by the GA, admit simple invariant
reductions for states with vorticities $\left( m_{1},m_{2}\right) =\left(
0,0\right) $ and $(1,1)$ or $\left( -1,-1\right) $, which correspond,
severally, to ansatz (\ref{ab}) with $a_{\pm }=b_{\pm }=0$, and $%
a_{0}=b_{0}=a_{-}=b_{-}=$ $0$ or $a_{0}=b_{0}=a_{+}=b_{+}=$ $0$. In
particular, for $\left( m_{1},m_{2}\right) =\left( 0,0\right) $ system (\ref%
{GalerkinSet}) reduces to a set of two equations:
\begin{eqnarray}
i\frac{da_{0}}{dt} &=&\left\vert a_{0}\right\vert ^{2}a_{0}-\kappa b_{0},
\notag \\
i\frac{db_{0}}{dt} &=&\left\vert b_{0}\right\vert ^{2}b_{0}-\kappa a_{0},
\label{two}
\end{eqnarray}%
with $\kappa >0$. This system with two degrees of freedom is integrable, as
it conserves Hamiltonian (\ref{H2}), written below, and quantities (\ref%
{integr}) (in this particular case, they amount to a single dynamical
invariant).

On the other hand, the set of dynamical variables which include vorticities $%
\left( 1,-1\right)$ or $%
\left( -1,1\right)$ {\ \emph{does not} correspond
to any invariant subsystem of Eq. (\ref{GalerkinSet}) (except for the
trivial case of }$\kappa =0${) with fewer than four degrees of freedom,
hence its evolution is governed by the} full system {(\ref{GalerkinSet}). We
use this system in all cases -- in particular, with the objective to test
stability of the invariant reductions with respect to small perturbations
which break the invariance. It is relevant to mention too that fixed points
of Eq. (\ref{GalerkinSet}) with }$\left\vert a_{+}\right\vert =\left\vert
a_{-}\right\vert $ and $\left\vert b_{+}\right\vert =\left\vert
b_{-}\right\vert $ may describe standing patterns of the wave functions $%
\sim \cos \left( m\theta \right) $ and/or $\sin \left( m\theta \right) $,
but we do not aim to address states of such types in the present work.

Thus, we simulated Eq. (\ref{GalerkinSet}) with different inputs,
corresponding to initial vorticity sets $\left( m_{1},m_{2}\right) =$ $(0,0)$%
, $(1,1)$, $(1,-1)$ and $(0,1)$, by means of the standard Runge-Kutta
algorithm. It was checked that the simulations indeed conserve the dynamical
invariants given by Eqs. (\ref{H}) and (\ref{integr}).

\begin{figure}[h]
\centering
\includegraphics[width=3.4in]{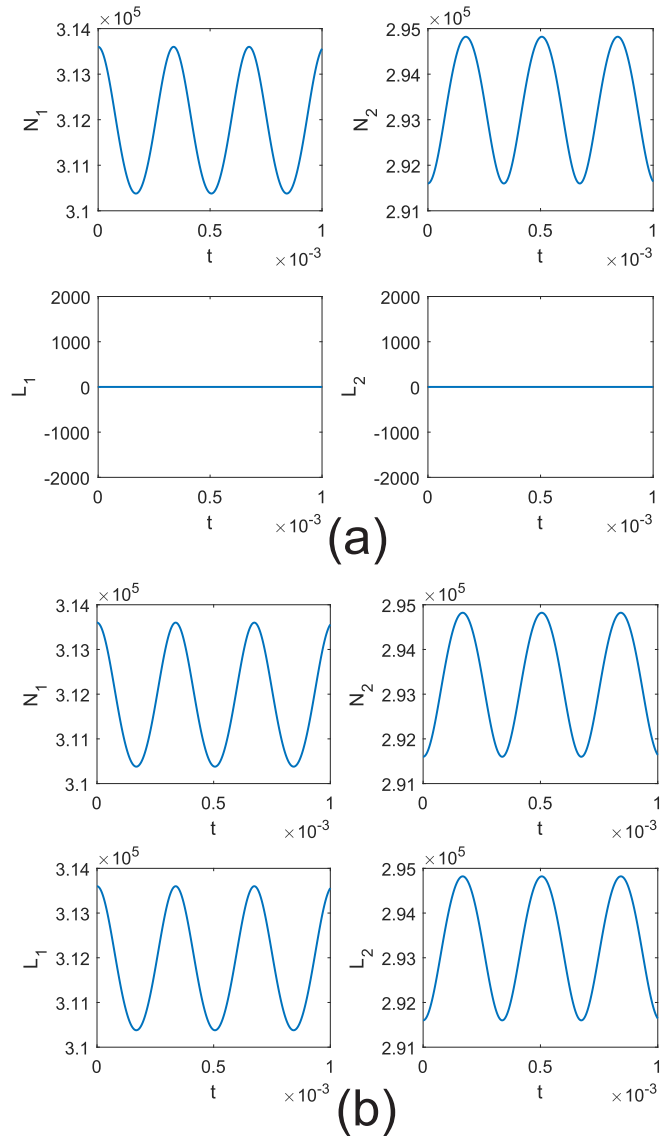}
\caption{The dynamics of the number of particles and angular momentum for
upper and lower rings with coupling $\protect\kappa =50$, as produced by
simulations of the GA system (\protect\ref{GalerkinSet}). The initial
conditions, which imply imbalance in the number of particles, are (a) $%
N_{1}=3.136\times 10^{5},N_{2}=2.916\times 10^{5}$, with vorticities $\left(
m_{1},m_{2}\right) =(0,0)$; (b) $N_{1}=3.136\times 10^{5},N_{2}=2.916\times
10^{5}$, with $\left( m_{1},m_{2}\right) =(1,1)$. }
\label{GalDyn(a)}
\end{figure}

\begin{figure}[h]
\centering
\includegraphics[width=3.4in]{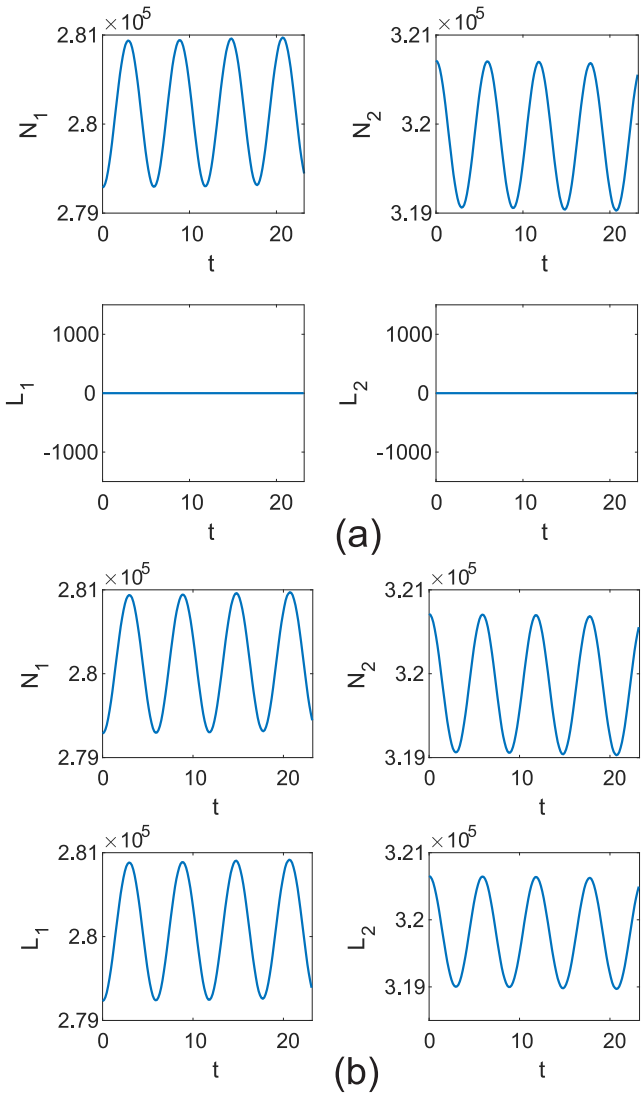}
\caption{The evolution of numbers of particles and angular momenta in the
top and bottom rings for the states generated by simulations of Eq. (\protect
\ref{GPE_dimless}) with inputs $\left( m_{1},m_{2}\right) =(0,0)$ (a); $(1,1)
$ (b). The scale of temporal variable in this figure is different from that
in its GA counterpart \protect\ref{GalDyn(a)}.}
\label{3DDyn(a)}
\end{figure}

\begin{figure}[h]
\centering
\includegraphics[width=3.4in]{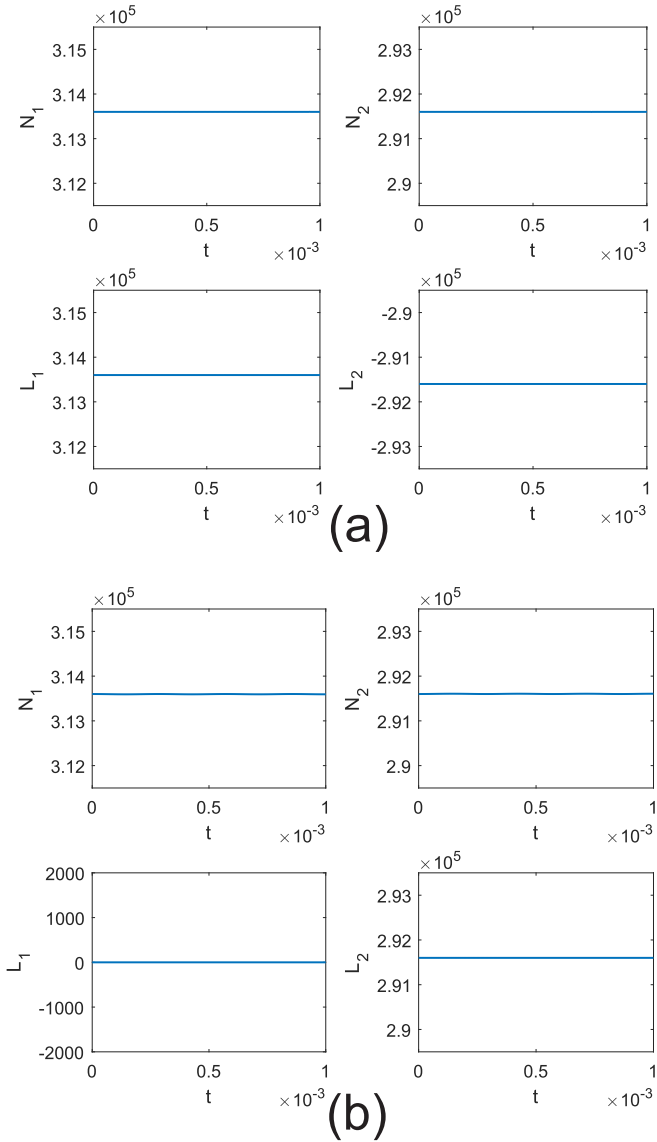}
\caption{The same as in Fig. \protect\ref{GalDyn(a)} for initial conditions
(a) $N_{1}=3.136\times 10^{5},N_{2}=2.916\times 10^{5}$, with vorticities $%
\left( m_{1},m_{2}\right) =$ $(1,-1)$; (b) $N_{1}=3.136\times
10^{5},N_{2}=2.916\times 10^{5}$, with $\left( m_{1},m_{2}\right) =(0,1)$.}
\label{GalDyn(b)}
\end{figure}

\begin{figure}[h]
\centering
\includegraphics[width=3.4in]{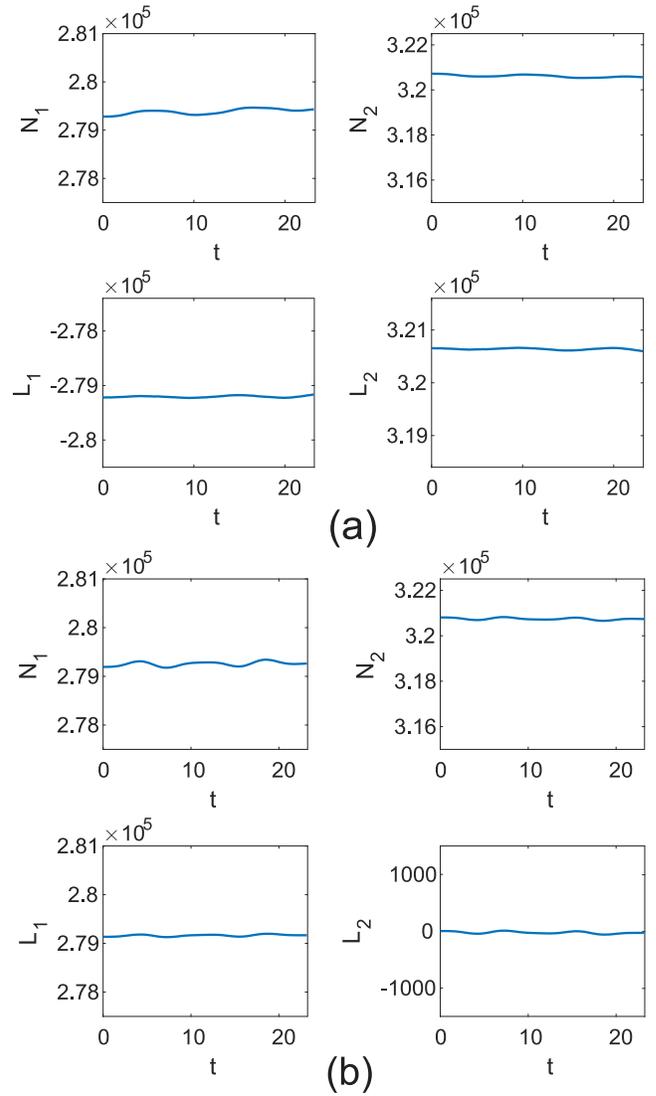}
\caption{The same as in Fig. \protect\ref{3DDyn(a)}, but for inputs $\left(
m_{1},m_{2}\right) =(-1,1)$ (a); $(1,0)$ (b).}
\label{3DDyn(b)}
\end{figure}

Figures \ref{GalDyn(a)} and \ref{GalDyn(b)} display the dynamics initiated
by inputs $(0,0)$, $(1,1)$, $(1,-1)$, and $(0,1)$ with a small initial
imbalance in the number of particles between the coupled rings, in the
strong-coupling regime ($\kappa =50$, which is responsible for small
oscillation period in this figure and similar ones). For the comparison's
sake, counterparts of these results, produced by full simulations of the 3D\
GPE (\ref{GPE3D}), are displayed close to them in Figs. \ref{3DDyn(a)} and %
\ref{3DDyn(b)}.\ Detailed discussion of the latter figures is given in the
next section.

In Figs. \ref{GalDyn(a)}(a) and \ref{GalDyn(a)}(b) the numbers of
particles oscillate similar to what is produced by the a.c. Josephson effect
in the double-well setting (as one can see in Figs. \ref{3DDyn(a)}(a) and \ref{3DDyn(a)}(b), we have obtained similar results by GPE). To support this conclusion, Fig. \ref%
{GalPhaseEvol00} demonstrates that the phase difference between the top and
bottom rings, $\delta \varphi =$arg$(\phi )$-arg$(\psi )$, for the input
with vorticities $\left( m_{1},m_{2}\right) =(0,0)$ input [for one of the $%
(1,1)$ type the situation is essentially the same] varies in time nearly
linearly, which is a characteristic feature of the a.c. Josephson effect
\cite{Levy07}. On the other hand, Figs. \ref{GalDyn(b)}(a,b) demonstrate
that the Josephson oscillations, as predicted by the GA, vanish for the
counter-rotating input, with $\left( m_{1},m_{2}\right) =(1,-1)$, as well as
for one of type $(0,1)$ (as one can see in Figs. \ref{3DDyn(b)}(a) and \ref{3DDyn(b)}(b) for $(-1,1)$, $(1,0)$ respectively, we have obtained similar results by GPE).

\begin{figure*}[thh]
\centering
\includegraphics[width=6.8in]{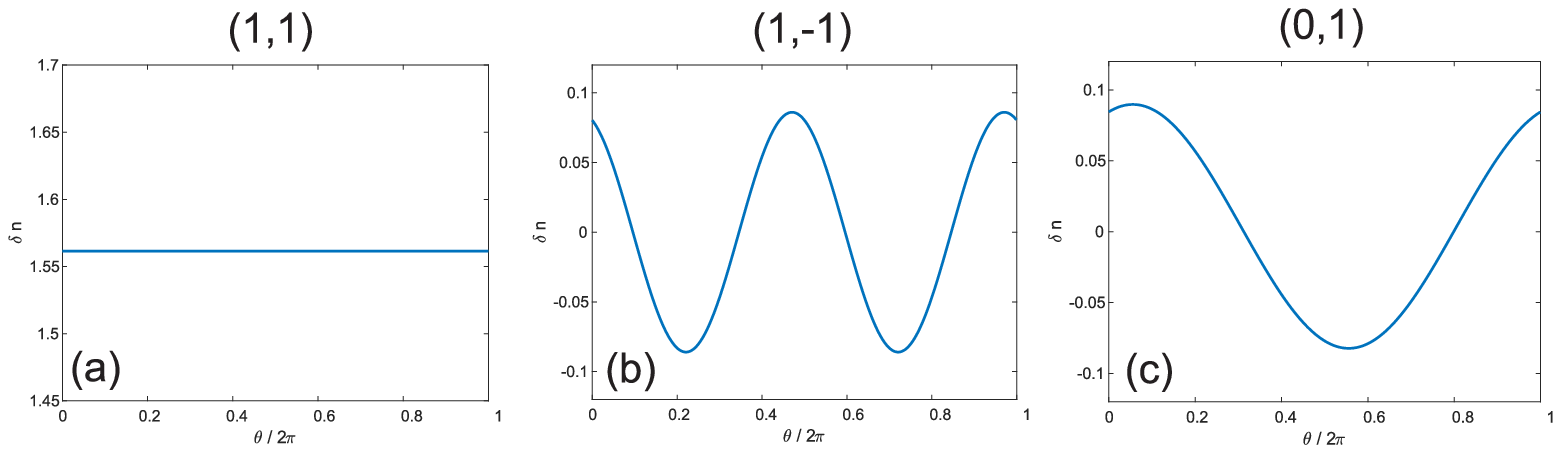}
\caption{Typical snapshots of the density variation (\protect\ref{delta_n}),
as predicted by the GA at $t=0.078\times 10^{-3}$ for the inputs with $%
\left( m_{1,}m_{2}\right) =$ $(1,1)$ (a); $(1,-1)$ (b); $(0,1)$ (c).}
\label{GalDensDistr}
\end{figure*}

It is also instructive to compare angular distributions of the inter-ring
tunneling flow. To this end, Fig. \ref{GalDensDistr} displays the density
variation,
\begin{equation}
\delta n_{\psi }(\theta ,t)\equiv \left\vert \psi (\theta ,t)\right\vert
^{2}-\left\vert \psi (\theta ,t=0)\right\vert ^{2},  \label{delta_n}
\end{equation}%
as a function of angular coordinate $\theta $. In particular, the state with
$\left( m_{1},m_{2}\right) =(1,1)$ does not develop the spatial variation,
while those of the $(1,-1)$ and $(0,1)$ types naturally build spatially
periodic patterns, with periods $T_{\theta }=\pi $ and $T_{\theta }=2\pi $,
respectively.

\begin{figure}[h]
\centering
\includegraphics[width=3.4in]{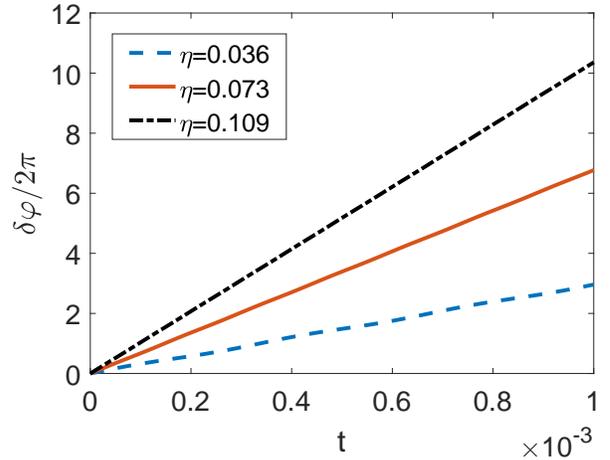}
\caption{(Color online) The evolution of the phase difference between the
wave functions in the upper and lower rings for the state generated by Eq. (%
\protect\ref{GalerkinSet}) with input $\left( m_{1},m_{2}\right) =(0,0)$ for
different values of imbalance $\protect\eta $ and fixed $\protect\kappa =50$.
}
\label{GalPhaseEvol00}
\end{figure}

\begin{figure}[h]
\centering
\includegraphics[width=3.4in]{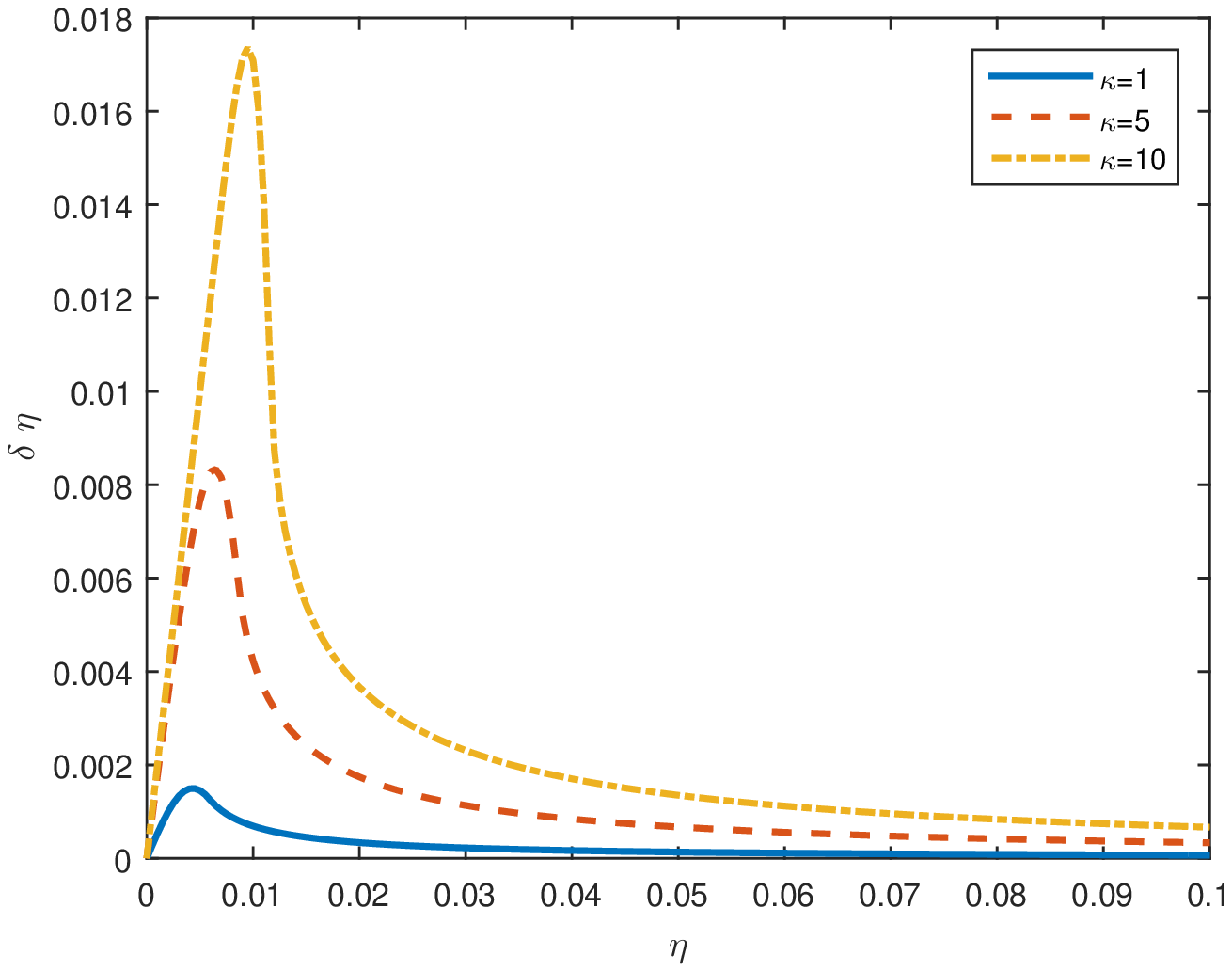}
\caption{(Color online) The GA-predicted asymmetry amplitude, $\protect%
\delta \protect\eta $, defined as per Eq. (\protect\ref{delta_eta}), as the
function of initial population imbalance $\protect\eta (t=0)$ for the state
with $\left( m_{1},m_{2}\right) =\left( 0,0\right) $ and different values of
coupling constant $\protect\kappa $. }
\label{delta_eta_vs_eta}
\end{figure}

The analysis of experimental data for the a.c. Josephson effect, which was
implemented in Ref. \cite{Levy07}, has demonstrated that the relative
difference in the number of particles,
\begin{equation}
\eta =(N_{1}-N_{2})/(N_{1}+N_{2}),  \label{eta=}
\end{equation}%
oscillates in time with a small amplitude $\sim \delta \eta \approx 10^{-2}$%
. %$\delta\eta\approx (1\div 1.14)\times 10^{-2}$.
It is relevant to consider restrictions on the amplitude of the oscillating
superflow in more detail. In the framework of the GA, the amplitude of
oscillations,%
\begin{equation}
\delta \eta =\eta _{\text{max}}-\eta _{\text{min}},  \label{delta_eta}
\end{equation}%
is determined by the initial value of the number-of-particle difference (\ref%
{eta=})\ and coupling constant $\kappa $. Figure \ref{delta_eta_vs_eta}
displays values of $\delta \eta $ averaged over ten oscillation periods for
state $\left( m_{1},m_{2}\right) =(0,0)$ with real initial conditions. As
expected, $\delta \eta $ grows with $\kappa $, and $\delta \eta \rightarrow 0
$ at $\eta (t=0)\rightarrow 0$. Note that the oscillation amplitude $\delta
\eta $ has a sharp maximum at small $\eta (t=0)$ and decays when the initial
asymmetry grows, due to the repulsive nonlinear interactions, which
resembles experimentally observed features of the macroscopic self-organized
oscillations of BEC trapped in a double-well potential \cite{Oberthaler}.

To gain further insight into the behavior of oscillation amplitude $\delta
\eta $, it is instructive to consider the simplest particular case, based on
the system of two ordinary differential equations (\ref{two}) for complex
amplitudes $a_{0}(t)$ and $b_{0}(t)$. The complex amplitudes can be
represented as%
\begin{equation}
a_{0}=|a_{0}|e^{i\alpha (t)},b_{0}=|b_{0}|e^{i\beta (t)},  \label{ab2}
\end{equation}%
where $|a_{0}|=\sqrt{A}\cos \left( \vartheta (t)\right) $, $|b_{0}|=\sqrt{A}%
\sin \left( \vartheta (t)\right) $, $A\equiv |a_{0}|^{2}+|b_{0}|^{2}$ is the
conserved total norm, and the distribution angle, $\vartheta $, varies in
interval $0\leq \vartheta \leq \pi /2$, so that $\sin \vartheta $, $\cos
\vartheta $ and norm $A$ are always positive (or zero).

The present system with two degrees of freedom is integrable (as mentiuoned
above), because it conserves $A$ and the Hamiltonian,%
\begin{gather}
H=\frac{1}{2}\left( |a_{0}|^{4}+|b_{0}|^{4}\right) -\kappa \left(
a_{0}b_{0}^{\ast }+a_{0}^{\ast }b\right)   \notag \\
\equiv \frac{1}{2}A^{2}\left[ 1-\frac{1}{2}\sin ^{2}\left( 2\vartheta
\right) \right] -\kappa A\sin \left( 2\vartheta \right) \cos \left( \alpha
-\beta \right) .  \label{H2}
\end{gather}%
In this case, asymmetry (\ref{eta=}) amounts to%
\begin{equation}
\eta \equiv \frac{\left\vert a_{0}\right\vert ^{2}-|b_{0}|^{2}}{%
|a_{0}|^{2}+|b_{0}|^{2}}\equiv \cos \left( 2\vartheta \right) ,  \label{eta}
\end{equation}%
hence the largest asymmetry corresponds to largest $\left\vert \cos
(2\vartheta )\right\vert $, i.e., smallest $\sin \left( 2\vartheta \right) $.

If the input corresponds to real initial values of $a_{0}(0)$ and $b_{0}(0)$
of the same sign [i.e., initially one has $\alpha _{0}=\beta _{0}=0$ in Eq. (%
\ref{ab2})], the value of $\sin \left( 2\vartheta \right) $ is determined by
the conservation of the Hamiltonian:%
\begin{gather}
\left[ \sin \left( 2\vartheta \right) \right] ^{2}+4\frac{\kappa }{A}\cos
\left( \alpha -\beta \right) \sin \left( 2\vartheta \right)   \notag \\
-\left[ \sin ^{2}\left( 2\vartheta _{0}\right) +4\frac{\kappa }{A}\sin
\left( 2\vartheta _{0}\right) \right] =0,  \label{quadr}
\end{gather}%
where $\vartheta _{0}$ is the initial value of $\vartheta $ [so that
constraint $\sin \left( 2\vartheta _{0}\right) \leq 1$ holds]. It is easy to
see that a local minimum (or maximum) of $\sin \left( 2\vartheta _{0}\right)
$, considered as a function of $\cos \left( \alpha -\beta \right) $, while
other parameters are fixed, does not exists. Indeed, differentiating Eq. (%
\ref{quadr}) with respect to $\cos \left( \alpha -\beta \right) $ and
setting $d\left( \sin \left( 2\vartheta \right) \right) /d\left( \cos \left(
\alpha -\beta \right) \right) =0$ shows that this condition may hold solely
at $\sin \left( 2\vartheta \right) =0$, but Eq. (\ref{quadr}) does have root
$\sin \left( 2\vartheta _{0}\right) =0$. Therefore, smallest and largest
values of $\sin \left( 2\vartheta \right) $ may only be attained at extreme
values of $\cos \left( \alpha -\beta \right) $, i.e.,$~\cos \left( \alpha
-\beta \right) =\pm 1$. The respective roots of Eq. (\ref{quadr}) are
\begin{eqnarray}
\sin (2\vartheta ) &=&\sin \left( 2\vartheta _{0}\right) ,  \label{=} \\
\sin \left( 2\vartheta \right)  &=&4\frac{\kappa }{A}+\sin \left( 2\vartheta
_{0}\right) .  \label{=4}
\end{eqnarray}%
This means that $\sin \left( 2\vartheta \right) $ never takes values smaller
than $\sin \left( 2\vartheta _{0}\right) $, hence the asymmetry cannot be
larger than its initial value. Note that, if $\kappa $ is a small parameter,
it follows from Eqs. (\ref{=}) and (\ref{=4}) that the amplitude of the
oscillations of the asymmetry may be approximated by
\begin{equation}
\delta \left( \cos \left( 2\vartheta \right) \right) \approx
4A^{-1}\left\vert \tan \left( 2\vartheta _{0}\right) \right\vert \kappa .
\label{delta}
\end{equation}%
This expression demonstrates that the amplitude of the asymmetry
oscillations is proportional to $\kappa $, and it has a sharp maximum at
small cos$\left( 2\vartheta _{0}\right) $, i.e., when the input has small
asymmetry. These features are in good agreement with numerical simulations
of Eq. (\ref{two}) for the state with vorticities $\left( m_{1},m_{2}\right)
=(0,0)$, see Fig. \ref{delta_eta_vs_eta}.

\begin{figure}[h]
\centering
\includegraphics[width=3.4in]{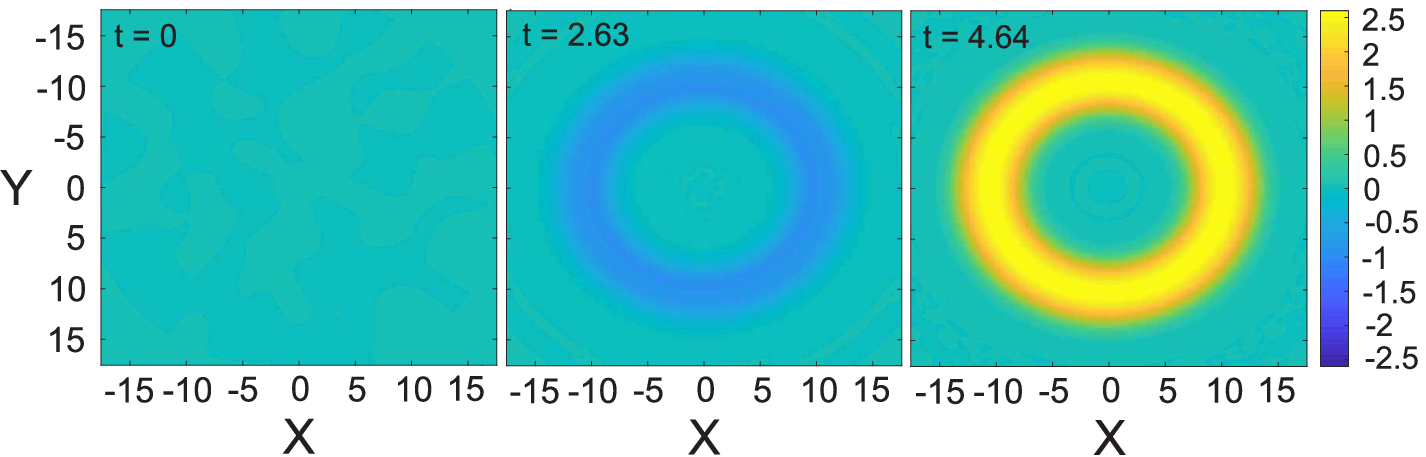}
\caption{(Color online) Snapshots at different moments of time of the
tunneling flow $j_{z}(x,y,z=0)$ for the state with $\left(
m_{1},m_{2}\right) =(1,1)$, as produced by simulations of Eq. (\protect\ref%
{GPE_dimless}).}
\label{PartFlow11}
\end{figure}

\begin{figure}[h]
\centering
\includegraphics[width=3.4in]{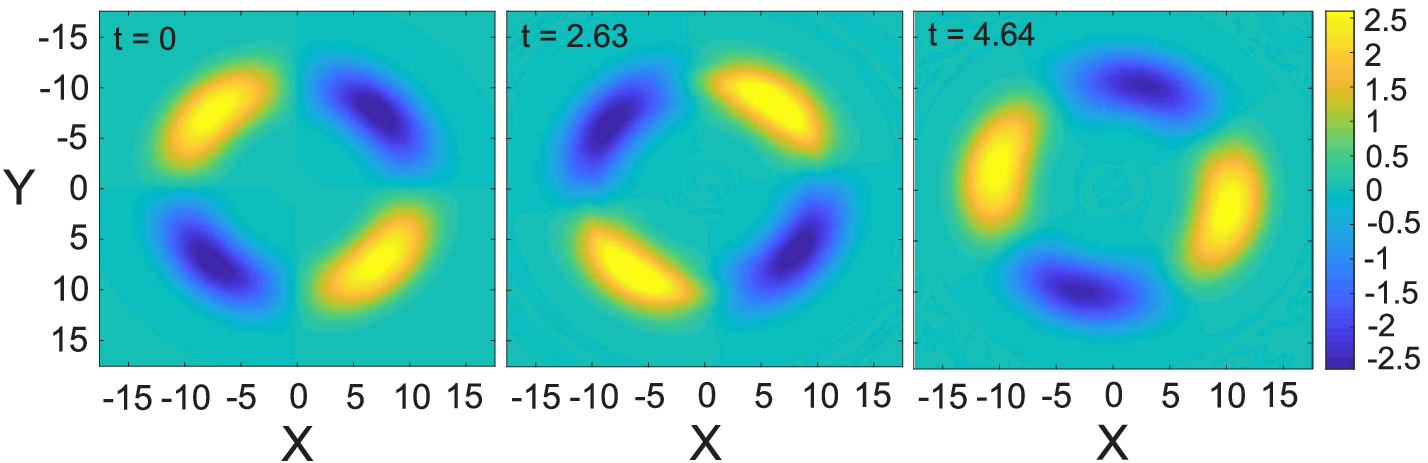}
\caption{(Color online) Snapshots at different moments of time of the
tunneling flow $j_{z}(x,y,z=0)$ for the state with $\left(
m_{1},m_{2}\right) =(1,-1)$, as produced by simulations of Eq. (\protect\ref%
{GPE_dimless}).}
\label{PartFlow1-1}
\end{figure}

\begin{figure}[h]
\centering
\includegraphics[width=3.4in]{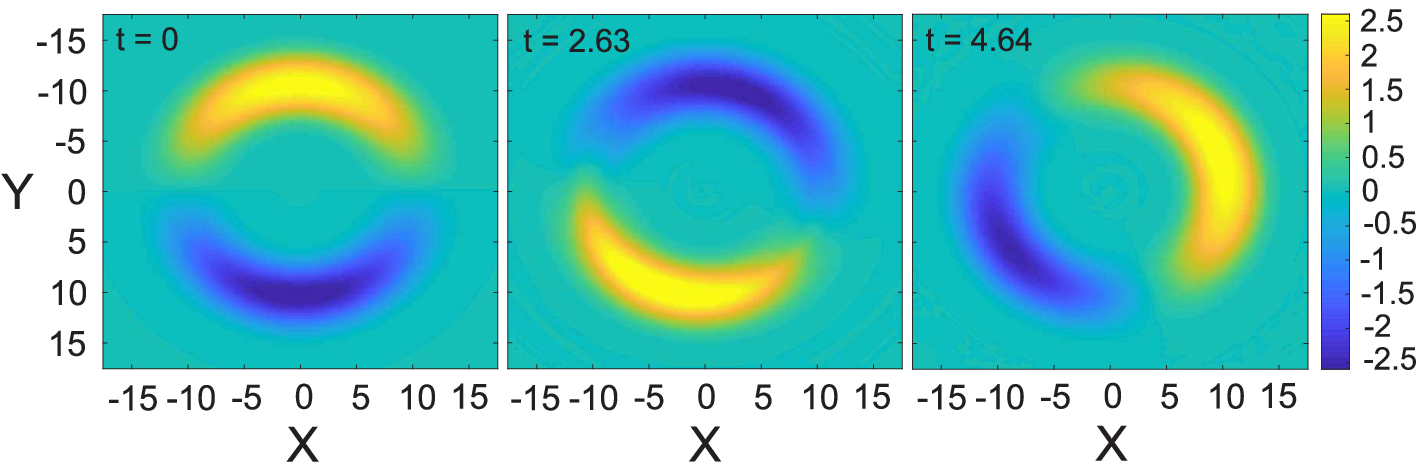}
\caption{(Color online) Snapshots at different moments of time of the
tunneling flow $j_{z}(x,y,z=0)$ for the state with $\left(
m_{1},m_{2}\right) =(0,1)$, as produced by simulations of Eq. (\protect\ref%
{GPE_dimless}).}
\label{PartFlow01}
\end{figure}

\begin{figure}[h]
\centering
\includegraphics[width=3.4in]{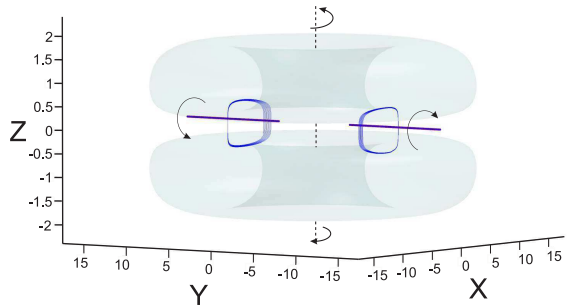}
\caption{(Color online) Vortex lines with vertically oriented cores
corresponding to the unbiased initial state with $\left( m_{1}, m_{2}\right)
=(-1,1)$, featuring a vortex-antivortex pair of fluxons with horizontally
oriented cores, as produced by simulations of Eq. (\protect\ref{GPE_dimless}%
). Superflow streamlines corresponding to the Josephson vortex-antivortex
pair are shown by blue lines.}
\label{3DCores}
\end{figure}

\begin{figure}[h]
\centering
\includegraphics[width=3.4in]{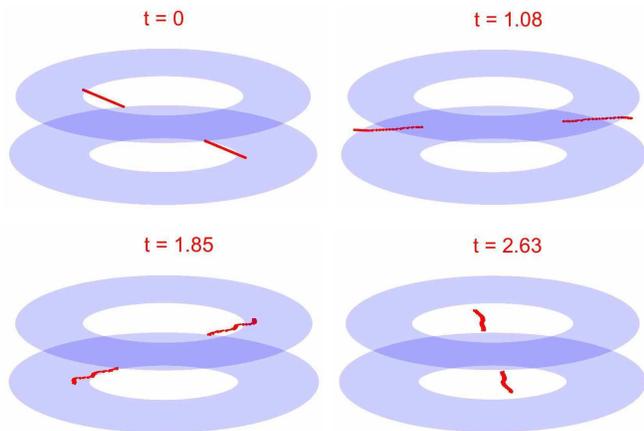}
\caption{(Color online) An isosurface of the double-ring condensate and
cores of the Josephson vortex and antivortex corresponding to the biased
initial state of the $(-1,1)$ type at different moments of time. Fluxon
cores rotate and bend, following the action of the quench with $U_{\protect%
\delta }=2\hbar \protect\omega _{r}$. }
\label{3DCoresBending}
\end{figure}

\section{Three-dimensional simulations}

In the framework of the underlying GPE (\ref{GPE3D}) with the external
potential taken as per Eq. (\ref{Vext}), the tunneling dynamics has been
simulated, assuming values of the physical parameters $\omega _{r}=2\pi
\times 123$ Hz, $\omega _{z}=2\pi \times 600$ Hz, $U_{b}=80\hbar \omega
_{r},a=0.3l_{r},l_{r}=\sqrt{\hbar /\left( M\omega _{r}\right) }=1.84$ $%
\mathrm{\mu }$m, $\rho _{0}=19.23$ $\mathrm{\mu }$m, which are appropriate
for the trapping pancake-shaped potential shown in Fig. \ref{BECRings}. The
number of atoms is fixed as $N=6\times 10^{5}$. Further, we define
dimensionless time, $t=\tau \omega _{r}$, and coordinates, $\mathbf{r}%
\rightarrow \mathbf{r}/l_{r}$, casting Eq. (\ref{GPE3D}) in the scaled form:
\begin{equation}
i\frac{\partial \Psi }{\partial t}=-\frac{1}{2}\nabla ^{2}\Psi +\widetilde{V}%
_{\mathrm{ext}}\Psi +\tilde{g}|\Psi |^{2}\Psi ,  \label{GPE_dimless}
\end{equation}%
with $\tilde{g}\equiv 4\pi a_{s}/l_{r}=0.0188$. Numerical solutions of the
GPE were used to calculate the superfluid-flow density as
\begin{equation}
\mathbf{j}(\mathbf{r})=\frac{i}{2}\left[ \Psi (\mathbf{r})\nabla \Psi
^{\ast }(\mathbf{r})-\Psi ^{\ast }(\mathbf{r})\nabla \Psi (\mathbf{r})\right]
.  \label{11}
\end{equation}

%We developed a 3D code to solve Eq. \ref{10}.
Stationary states of the form $\Psi (\mathbf{r},t)=\Psi (\mathbf{r})e^{-i\mu
t}$, with chemical potential $\mu $, can be found numerically by means of
the imaginary-time-propagation method \cite{im-time}. To obtain a stationary
state with different vortex phase profiles in the upper and lower rings, we
initiate the evolution in imaginary time with following input:
\begin{equation}
\Psi (\mathbf{r})=|\Psi _{0}(x,y,z)|e^{iS(z)\theta },  \label{input}
\end{equation}%
where $\Psi _{0}(x,y,z)$ is a numerically found solution with zero vorticity
in both rings, $\theta $ is, as
above, the polar angle in the cylindrical coordinates, and integer
topological charges $m_{1}$ and $m_{2}$ are imprinted in accordance with
angular momenta carried by the top and bottom rings: $S(z)=m_{1}$ for $z\geq
0$ and $S(z)=m_{2}$ for $z<0$, cf. a similar procedure used with a
``peanut-shaped" trapping potential developed in Ref. \cite%
{hybrid}.

The dynamics of BEC in real time was simulated by means of the usual
split-step fast-Fourier-transform method. In the simulations, a difference
between the top and bottom condensates was seeded by a population imbalance
in the initial state of the double-ring system. First, we find a stationary
state in an asymmetric setting, with the potential of the bottom ring, at $%
z<0$, biased by a constant term $U_{\delta }>0$. Then, at $t=0$ the biased
state is quenched by suddenly turning the asymmetry off, $U_{\delta
}\rightarrow 0$, which was followed by real-time simulations of Eq. (\ref%
{GPE_dimless}) at $t>0$. Each value of $U_{\delta }$ introduces a specific
value of the chemical-potential difference between the top and bottom
condensates. We use $U_{\delta }=1.1\hbar \omega _{r}$ for the simulations
reported below, which is a physically relevant value of the bias. We have
also simulated the evolution of asymmetrically quenched inputs in other
forms (for example, using a 3D analog of a linearly-tilted double-well
potential). It was found that the particular form of the initial asymmetry
does not essentially affect the ensuing dynamics of the tunneling flows.

In the experiment, the double-ring system with different angular momenta in
its top and bottom parts may appear spontaneously as a result of cooling,
with different momenta, $m_{1}$ and $m_{2}$, being frozen into the two rings
after the transition into the BEC state. Such hybrid states can also be
created in a more controllable way. Indeed, the asymmetry of the density
distribution in the top and bottom rings makes it possible to excite the
vorticity by applying a stirring laser beam to one ring only, keeping the
mate one in the zero-vorticity state \cite{Wright2013, PRA15_2}. Thus one
can prepare the input in the form of the $(0,1)$ or $(1,0)$
state. Further, using the phase-imprinting technique, it is possible to
create a $(-1,1)$ or $(1,-1)$ state with the help of a properly tuned Laguerre-Gauss
beams LG$_{-1,0}$, which resonantly interacts with atoms in zero-vorticity
state.

As well as the simplified 1D model introduced above in the form of Eq. (\ref%
{coupled}), the 3D GPE conserves the Hamiltonian, total number of atoms (in
the scaled form), $\int |\Psi |^{2}d\mathbf{r}=N$, and the total angular
momentum, $L_{z}$. Naturally, the evolution of the atomic populations and
angular momenta in the bottom and top rings crucially depends on vorticities
$m_{1}$ and $m_{2}$ in the hybrid state. In particular, one can observe
periodic oscillations of the populations in the states of types $(m_{1}$, $%
m_{2})=(0,0)$ and $(1,1)$, as seen in Fig. \ref{3DDyn(a)}(a) and (b). The
oscillation amplitude is determined by initial population differences, as it
was pointed out above, using the GA (see Fig. \ref{delta_eta_vs_eta}). It
was also mentioned above that BEC trapped in the double-well potential, with
a tunnel link connecting the wells, features Josephson oscillations between
them \cite{Levy07,fnt18}. In our numerical simulations we have obtained a
linear time dependence of the phase difference between the top and bottom
rings for the states with $m_{1}=m_{2}$ (see Supplemental Material), similar
to what was produced by the GA, see Fig. \ref{GalPhaseEvol00}. The linear
time dependence corroborates that the oscillating atomic flow may be
understood as the a.c. Josephson regime. Further, Figs. \ref{3DDyn(b)}(a,b)
demonstrate that the inputs with vorticity sets $\left( m_{1},m_{2}\right)
=\left( -1,1\right) $ and $\left( 1,0\right) $ generate no Josephson
oscillations, again in agreement with the GA predictions, cf.  Fig. \ref{3DDyn(b)}.

The spatial structure of the patterns of different types are adequately
illustrated by snapshots of the tunneling flow through the barrier between
the rings, $j_{z}(x,y,z=0)$, which is defined by\ Eq. (\ref{11}). A set of
such snapshots, taken at different moments of time, which are produced by
simulations of Eq. (\ref{GPE_dimless}), are displayed in Figs. \ref%
{PartFlow11}, \ref{PartFlow1-1} and \ref{PartFlow01}. The angular
distributions of $j_{z}(x,y,z=0)$ in the Josephson junctions in patterns of
all the types considered here are in excellent agreement with the
predictions of the GA, see Figs. \ref{PartFlow11}-\ref{PartFlow01} and \ref%
{GalDensDistr}.

Thus, the numerical simulations of the GA and full GPE predict the same
effects in the superflow dynamics in the double ring: in the non-rotating $%
(0,0)$ and co-rotating $(1,1)$ states one observes generic Josephson
oscillations of the total tunneling flow, while for the hybrid
counter-rotating state of the $(1,-1)$ and $(-1,1)$ type, as well as for the semi-vortex
one, with topological charges $(0,1)$ and $(0,1)$, the total tunneling flow
vanishes, which resembles an effect known for cylindrical superconductive
Josephson junctions \cite{Sherill79,Sherill81}.

To gain a deeper insight into the hybrid dynamical state, produced by the
input with different vorticities in the two rings, one may again use the
superfluid current (\ref{11}). To analyze its structure, we substitute
ansatz (\ref{3}) for the 3D wave function, with $\Phi _{+}=|\Phi
_{+}|e^{-i\mu _{+}t},\Phi _{-}=|\Phi _{-}|e^{-i\mu _{-}t}$ (wave functions
of the stationary states), and $\phi =e^{im_{1}\theta },\psi
=e^{im_{2}\theta }$. The substitution yields
\begin{equation}
j_{z}\sim \sin [(\mu _{+}-\mu _{-})t+(m_{2}-m_{1})\theta ].  \label{18}
\end{equation}%
This simple relation agrees well with full 3D simulations, see Figs. \ref%
{PartFlow1-1} and Fig. \ref{PartFlow01}, as well as with the GA predictions,
see Figs. \ref{GalDensDistr} (b,c)). In particular, it follows from Eq. (\ref%
{18}) that the total inter-ring currents for the states of the $(1,-1)$ and $%
(0,1)$ types indeed vanish:
\begin{equation}
J_{\mathrm{total}}=\int_{0}^{2\pi }j_{z}d\theta =0,  \label{19}
\end{equation}%
as found above from the 3D simulations [Figs. \ref{3DDyn(b)} (a,b)] for $(-1,1)$ and $%
(1,0)$ and GA
[Figs.\ref{GalDyn(b)} (a,b)] for $(1,-1)$ and $%
(0,1)$.

On the other hand, for the states of the $\left( m_{1},m_{2}\right) =(1,1)$
and $\left( 0.0\right) $ types [$\phi =e^{i\theta },\psi =e^{i\theta }$, or $%
\phi =1,\psi =1$, respectively in Eq. (\ref{3})], Eqs. (\ref{11}) and (\ref%
{19}) produce a nonvanishing a.c. Josephson effect: $J_{\mathrm{total}}\sim
\sin [(\mu _{+}-\mu _{-})t].$ This conclusion is again in agreement with the
results of both the 3D simulations [Figs. \ref{3DDyn(a)} (a,b) and \ref%
{PartFlow11}] and the GA prediction [Fig. \ref{GalDyn(a)} (a,b) and \ref%
{GalDensDistr}(a)]. Thus we conclude that, in the general case, the states
of the type $(m_{1},m_{2})$ with $m_{1}=m_{2}$ feature the a.c. Josephson
effect, while ones with $m_{1}\neq m_{2}$ produce zero total current.

%\section{Fluxons}

Finally, angular distributions of the tunneling superflow $j_{z}$ for hybrid
states of the $(-1,1)$ and $(1,0)$ types are displayed in Figs. \ref%
{PartFlow1-1} and \ref{PartFlow01}. The structure of the tunneling flows
suggests formation of a Josephson vortex (fluxon) in the $(1,0)$ state, and
of a vortex-antivortex pair for the $(-1,1)$ configuration, see Fig. \ref%
{3DCores}. For the symmetric unbiased initial state, one indeed observes
stationary Josephson vortices. This result is similar to findings reported
in Ref. \cite{Brand18}, in which stationary vortices were obtained in an
array of linearly-coupled one-dimensional Bose-Einstein condensate. However,
following the application of the quench, in the biased two-ring system we
observe, in Fig. \ref{3DCoresBending}, that fluxon cores rotate and bend
(see also \cite{SupMat}). It is easy to see from Eq. (\ref{18}) that the
fluxon's azimuthal cycling frequency is determined by chemical-potential
difference. These predictions of Eq. (\ref{18}), obtained by the GA, are
found to be in excellent agreement with extensive series of numerical
simulations for different values of the initial population unbalance.
%%%%%%%%%%%%%%%%%%%
%\newpage

\section{Conclusion and discussion}

We have considered the bosonic Josephson junction between two atomic
condensates loaded in parallel-coupled ring-shaped traps. The analysis is
performed using the truncated model produced by the GA (Galerkin
approximation), and through direct systematic simulations of the underlying
three-dimensional GPE. The approximation reducing the full GPE to a system
of linearly coupled 1D equations is employed too. These approaches
demonstrate that the a.c. Josephson effect with a uniform angular
distribution of the superflow tunneling between the rings can be observed in
states with equal vorticities in the parallel-coupled rings, $m_{1}=m_{2}$.
In hybrid states with different vorticities ($m_{1}\neq m_{2}$),
%; in fact, we have considered ones of the $\left( 0,1\right)$
%and $\left( 1,-1\right) $ types]
the total flow of the inter-ring tunneling vanishes, the angular
distribution of the tunneling superflow being a periodic function of the
angular coordinate with zero average. It is remarkable that the vanishing of
the net tunneling superflow is accompanied by the formation of $%
N_{J}=|m_{1}-m_{2}|$ Josephson vortices (fluxons) trapped in the Bose
Josephson junction. Detailed consideration of interactions between the
fluxons in a regime of strong ring-ring coupling may be a relevant extension
of the present work. Another relevant direction for the continuation of the
analysis may be consideration of two-layer settings for spinor
(two-component) condensates \cite{spinor1,spinor2}.

\section*{ACKNOWLEDGMENTS}

The work of B.A.M. is supported, in a part, by the Israel Science
Foundation, through grant No. 1287/17. A.Y. acknowledges support from
Project "Topological properties of chiral materials and Bose-Einstein
condensates in magnetic field" by Ministry of Science and Education of
Ukraine.

\newpage

\end{document}